\documentclass{elsart}

\usepackage{natbib}

 \usepackage{graphicx}

\usepackage{amssymb}

\begin{document}

%
\def\etal{et {\it al.} }
\def\araa{{\it Ann.\ Rev.\ Astron.\ Ap.}}
\def\aplet{{\it Ap.\ Letters}}
\def\aj{{\it Astron.\ J.}}
\def\apj{ApJ}
\def\apjl{{\it ApJ\ (Lett.)}}
\def\apjs{{\it ApJ\ Suppl.}}
\def\aas{{\it Astron.\ Astrophys.\ Suppl.}}
\def\aa{{\it A\&A}}
\def\aap{{\it A\&A}}
\def\mnras{{\it MNRAS}}
\def\nat{{\it Nature}}
\def\pasa{{\it Proc.\ Astr.\ Soc.\ Aust.}}
\def\pasp{{\it P.\ A.\ S.\ P.}}
\def\pasj{{\it PASJ}}
\def\pre{{\it Preprint}}
\def\sovlet{{\it Sov. Astron. Lett.}}
\def\adspr{{\it Adv. Space. Res.}}
\def\expas{{\it Experimental Astron.}}
\def\ssr{{\it Space Sci. Rev.}}
\def\apss{{\it Astrophys. and Space Sci.}}
\def\inpress{in press.}
\def\souspresse{sous presse.}
\def\inprep{in preparation.}
\def\enprep{en pr\'eparation.}
\def\submit{submitted.}
\def\soumis{soumis.}
\def\aph{{\it Astro-ph}}
\def\astroph{{\it Astro-ph}}

\def\xmm{XMM-{\it Newton}}


\newcommand{\hms}[3]{#1$^h$#2$^m$#3$^s$}
\newcommand{\dms}[3]{#1$^\circ$#2$'$#3$''$}

\begin{frontmatter}



  \title{Probing the dark matter profile of hot clusters and the $M-T$
    relation with \xmm}


\author[label1]{E. Pointecouteau,}
\author[label1]{M. Arnaud,}
\author[label2]{G.W. Pratt}

\address[label1]{SAp/CEA, l'Orme des Merisiers, 91191 Gif-sur-Yvette, France}
\address[label2]{MPE, Giessenbachstra{\ss}e, 85748 Garching, Germany}

\begin{abstract}
  
  We present results based on \xmm~ observations of a small sample of
  hot galaxy clusters. Making a full use of \xmm's spectro-imaging
  capabilities, we have extracted the radial temperature profile and
  gas density profile, and with this information, calculated the total
  mass profile of each cluster (under the assumption of hydrostatic
  equilibrium and spherical symmetry). Comparing the individual scaled
  total mass profiles, we have probed the Universality of rich cluster mass
  profiles over a wide range of radii (from 0.01 to 0.7 the virial
  radius). We have also tested the shape of cluster mass profiles by
  comparing with the predicted profiles from numerical simulations of
  hierarchical structure formation. 
  We also derived the local mass-temperature ($M-T$) scaling relation
  over a range of temperature going from 4 to 9~keV, that we compare
  with theoretical predictions.

\end{abstract}


\end{frontmatter}

\section{Introduction}
\label{intro}

In a framework where cosmic structure forms under the effect of
gravitation alone, the population of galaxy clusters should be
strongly similar.  A similarity of shape is expected for the dark
matter distribution in clusters due to the process of gravitational
collapse.  The gas ``follows'' the distribution of the dominant dark
matter component within the gravitational well, and is in approximate
hydrostatic equilibrium (between major mergers). Then the matter in
clusters (be it baryonic or dark) has a similar distribution from
cluster to cluster. Scaling laws are another aspect of cluster
similarity and are a direct prediction from scenarios of structure
formation based solely on gravitation.  They express the correlation
between physical quantities (such as temperature, luminosity,
entropy,...), and in fact make clusters a two parameter population
which depends only on the redshift and the total mass of each object.
Each single other quantity can then be expressed as follows:
$Q(z)=a(z)M^\alpha$.

The properties of cluster similarity are a powerful statistical tool
for the study of the cluster population, and a rich source of
information on structure formation and evolution. (For details see
\citet{arnaud04} and reference therein).

The paper presents results in the currently-favored concordance
cosmological model: $H_0=70$~km/s/Mpc, $\Omega_m=0.3$ and
$\Omega_\Lambda=0.7$.

\section{Universality of the mass distribution}
\label{univ}

We have investigated the shape of the dark matter profile in clusters
of galaxies using a sample of five rich, relaxed nearby clusters with
temperatures ranging between 4 and 9~keV. The different steps in the
computation of each mass profile are:

\begin{itemize}
\item The computation of an azimuthally-averaged surface brightness
(SB) profile, which is then parameterized with either (a) the sum of
$n$ $\beta$-models ($n=2 \textrm{ or } 3$, BB or BBB model, \citealt{pointecouteau04}) or (b) a modified double $\beta$-model allowing a peaked density distribution toward the cluster center (i.e KBB model, \citealt{pratt02}).

\item Spatially-resolved spectroscopic fits of annular regions to obtain a
temperature profile. The output projected temperature profile is then
corrected for projection effects and for the effects of PSF
blurring. This is achieved through a Monte Carlo simulation in which
the projected profile is randomly sampled 1000 times within the
observed errors, parameterized using the functional form of
\citet{allen01}, and subsequently corrected for projection and PSF
effects.

\item Under the assumptions of spherical symmetry and hydrostatic
equilibrium, these density and the temperature profiles can then be
used to derive the total mass profile.  More detailed explanations of
those different steps can be found in
\citet{pratt02,pratt03,pratt04} and \citet{pointecouteau04}.

\end{itemize}

 \begin{figure}
\includegraphics[width=13cm]{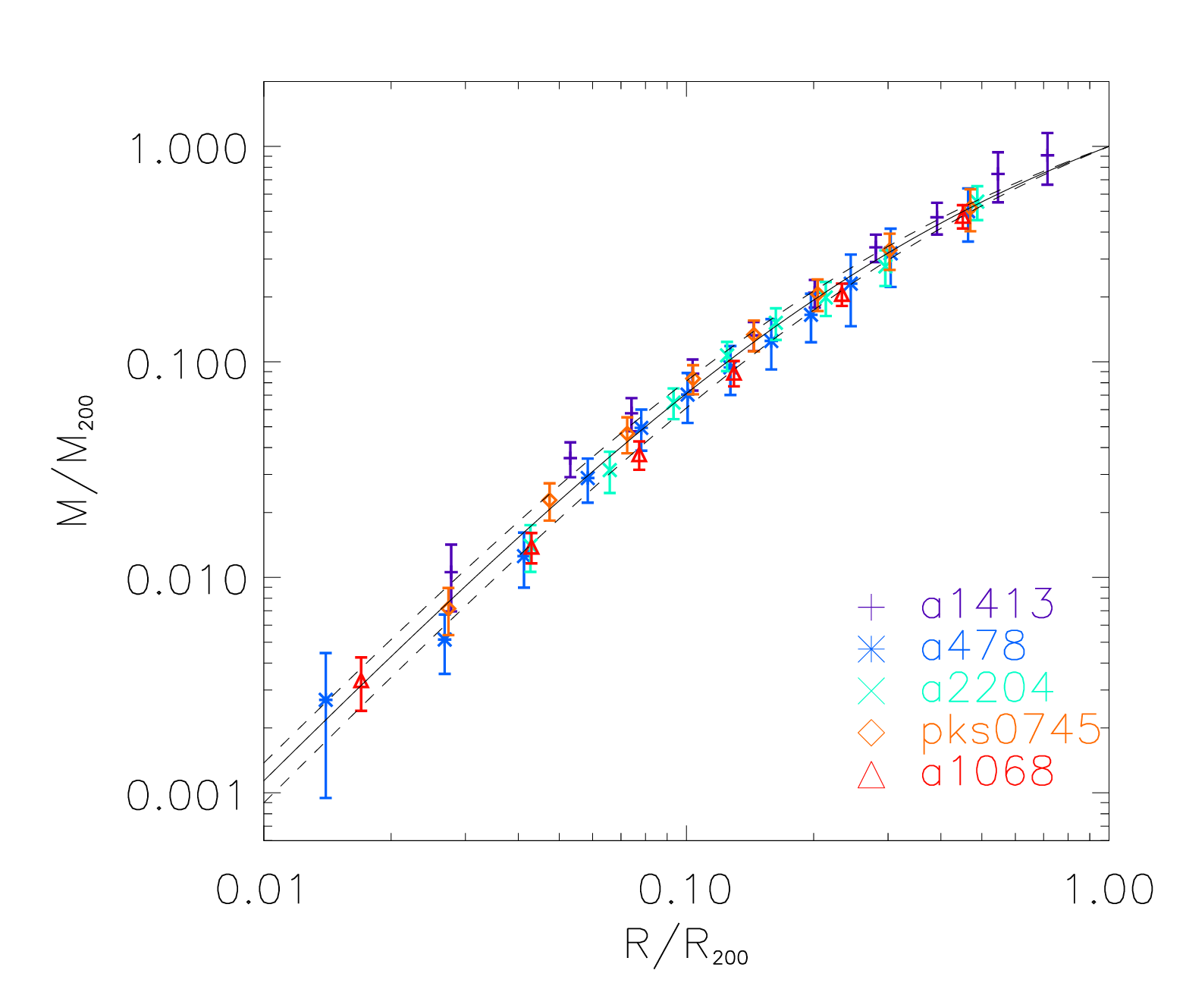}
 \caption{Scaled mass profiles. Each profile is scaled with
   respect to its virial radius $R_{200}$ and virial mass $M_{200}$
   (see text).  The solid black line corresponds to the average NFW
   profile and the two dashed lines are the associated standard
   deviation.}
 \label{fig1}
\hspace*{0.5cm}\includegraphics[width=12cm]{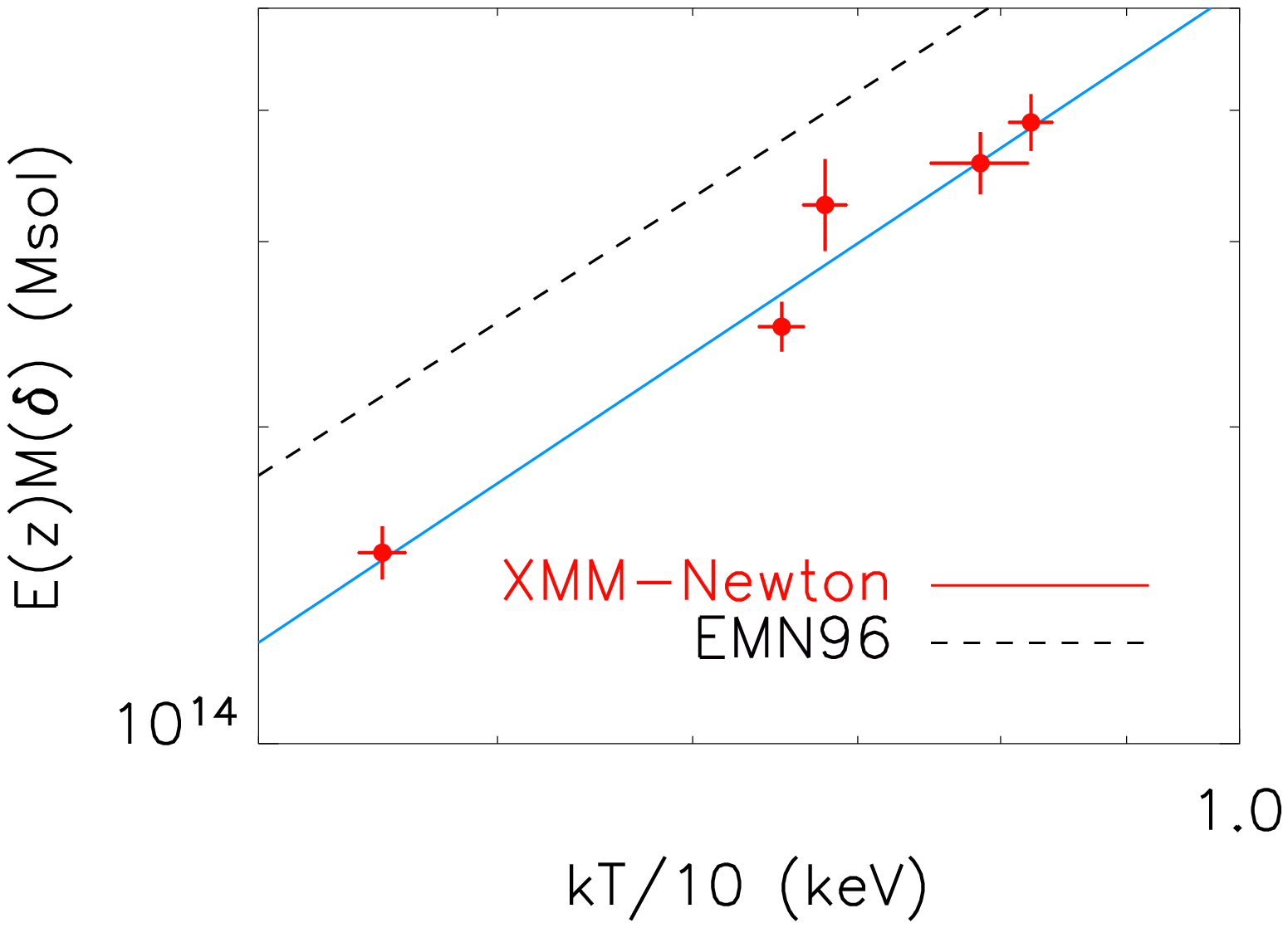}
 \caption{The $M-T$ relation at $\delta=2500$ from 5
   hot relaxed clusters observed with \xmm\ (red filled dots with
   associated error bars); the associated best fit is overplotted as a
   solid red line. 
   The dashed black line is the prediction
     from numerical simulations based purely on gravitational
     structure formation \cite{emn96}.}
 \label{fig2}

 \end{figure}

We have used our derived mass profiles to test different models for
the dark matter distribution. We tested an isothermal sphere model, a
MQGSL profile \citep{mqgsl99} and a NFW profile \citep{nfw97}. The NFW
profile proved to be the best representation of our current data. We
adopted this as the parametric description for our cluster mass
profiles. In fitting the mass profiles, the free parameters were the
mass at a given overdensity with respect to the critical
density of the Universe, $M(\delta)$, and either the scale radius,
$R_s$, or the concentration parameter, $c$. This approach allows us to
work directly with the mass at a chosen overdensity. Each pair of
parameters led to identical results. 

We then scaled each total mass profile according to the cluster virial
radius (defined as the radius where the density is equal to 200 times
the critical density of the Universe at the cluster redshift,
$R_{200}$) and the virial mass (the mass at this radius, $M_{200}$).
Our small sample includes the following clusters: A478, PKS0745, A1068
, A1413 and A2204.  The results on A1413 and A478 have been
respectively published by \citet{pratt02} and \citet{pointecouteau04}.
PKS0745 was originally analyzed by \citet{chen03}, but we reprocessed
the whole data set checking that our result are consistent with the
previously published ones. Nothing on A1068 and A2204 has yet been
published. The details on clusters are gathered in
Table~\ref{tab:tab1}).  
The five scaled mass profiles are plotted
together in Fig.~\ref{fig1}.  The five profiles scale very well over a
large range of radii from about 0.01 to 0.7 the virial radius,
exhibiting a strongly similar shape as expected for an underlying
universal matter profile in clusters.

\begin{table}
\caption{Details on the sample \label{tab:tab1}}
\begin{center}
\begin{tabular}{lcccc}
\hline
Name & $z$ & Rev. & $t_{exp}$ & $n_e(r)$ model \\
\hline
A478 & 0.088 & 401,411& 48/41/37& BBB\\
PKS0745 & 0.103& 164& 10/10/--& KBB\\
A1068 & 0.137& 633& 19/20/15& BBB\\
A1413 & 0.143& 182& 24/25/10& KBB\\
A2204 & 0.152& 322& 20/20/13& KBB\\
\hline
\end{tabular}
\end{center}
~\\[0.5em]
{\footnotesize
Columns: (1) Name. (2) redshift. (3) XMM-Newton revolution. (4)
Effective exposure time for EMOS1/EMOS2/EPN. (4) Parametric model used
for the density profile modelisation.}
\end{table}

\section{The $M-T$ relation as seen by \xmm}
\label{mt}

In order to further check the properties of cluster similarity with
our sample, we have investigated the mass to temperature ($M-T$)
relation. For easy comparison with previous work, we focus on an
overdensity of $\delta=2500$ with respect to the critical density of
the Universe.  The ``virial'' temperatures used to determine the $M-T$
relation were obtained from direct spectroscopic fits over a region
from $r>0.1 R_v$, chosen to avoid the cooling cores found in many of
these clusters.  We then performed a linear regression fit in the
$\textrm{log}\,M-\textrm{log}\,kT$ plane, taking into account both the
errors in $M$ and $kT$. With only five \xmm~ clusters we are able to
derive the following $M-T$ relation: $M_{2500}=1.45\times
10^{13}(kT/\textrm{keV})^{1.56\pm 0.16}$~M$_\odot$ (see Fig.~\ref{fig2}). This
result is compatible in terms of slope with the slope of 1.5 predicted
by numerical simulations relying on gravitation-based scenarios of
structure formation \citep{emn96}. However, the normalization of the
observed relation is lower by a factor of about 30\%, a problem
encountered in several previous works
\citep{nevalainen00,finoguenov01,allen01,sanderson03}.  We then added
the published {\it Chandra\/} data on four rich ($kT > 5$ keV)
clusters not already in our sample (A2390, A1835, RXJ1347-1145,
MS2137-2353; \citealt{allen01}) and re-derived the $M-T$ relation,
finding $M_{2500}=1.46\times 10^{13}(kT/\textrm{keV})^{1.55\pm
  0.15}$~M$_\odot$. The fit is almost unchanged, perhaps due to the
larger uncertainties of the Chandra measurements with respect to the
very precise \xmm~ measurements.

 Thus at $\delta=2500$, our derived $M-T$ relation is compatible in
 terms of slope with the prediction from scenarios of structure
 formation based on gravitation alone.

\section{Conclusion}
\label{conc}

With this work we have used \xmm's excellent imaging and
spectroscopic capabilities to observe clusters of galaxies up to about
half of the virial radius ($r\sim0.5R_{200}$). In this study we were able
to:

\begin{enumerate}
\item probe the universality of the dark matter distribution in a
sample of 5 rich, nearby galaxy clusters from 0.01 to 0.7 times the
virial radius. The five scaled mass profiles present a very similar
shape which is fairly well described by an NFW density distribution.

\item investigate the $M-T$ relation at the overdensity of 2500 times
  the critical density of the Universe over a range of temperature
  going from about 4~keV to 9~keV. The derived $M-T$ relation has a
  slope compatible with the expected slope of 1.5 from the numerical
  simulations based on gravitational heating. The precision of the
  \xmm\ measurements allows us to put tight constraints on the $M-T$
  slope with only 5 clusters. Still our observed normalization for
  this relation is lower by a factor of about 30\% compared to
  theoretical predictions.

\end{enumerate}

Since the mass is not measured directly, but is derived from the
density and the temperature profiles, a key issue for this kind of
work lies in the accuracy of the determination of the true (meaning
deprojected and PSF corrected) temperature and surface brightness
profiles. The shape of Chandra temperature profiles (Vikhlinin, this
meeting) and the \xmm\ profiles we derived from our observations have
therefore to be cross-compared. This is essential to assess the
reliability of temperature profiles in mass profile determination and
in the calibration of scaling laws. 
In a forthcoming paper, we extend our sample coverage to
lower temperature (Pointecouteau, Arnaud \& Pratt, in preparation). There we
will investigate in greater detail the properties of the scaled profiles,
the slope and normalization of the $M-T$ relation, and compare our results
to recent numerical simulations involving non-gravitational processes.

The quality of current X-ray data makes it possible to calibrate the
scaling laws in the local universe (the highest redshift cluster of
our sample is A2204 with $z=0.1523$) with a high precision which is
fundamental for: (i) further studies concerning the evolution of
scaling laws, and therefore for the studies of structure formation and
evolution; (ii) surveys at different wavelengths like blind SZ
surveys, which will make use of calibrated X-ray scaling laws as a
bridge to calibrate SZ scaling laws such as the $Y-M$ relation
(roughly the SZ flux to total mass relation), needed for cosmological
studies based on SZ statistics.

Indeed for such kind of studies, biases and/or systematic errors could
be induced by the thermodynamical state of the clusters, or by some
hypothesis or processes in the data analysis. Such kind of biases and
systematics have to be carefully investigated and quantified to assess
the calibration of scaling laws, and to make their use reliable on
large cluster samples for cosmological studies and structure formation
studies.

\end{document}